# PUBLIC KEY PROTOCOL BASED ON AMALGAMATED FREE PRODUCT[*]

SUMIT KUMAR UPADHYAY[1], SHIV DATT KUMAR[2]
RAMJI LAL[3]

[1,2]MATHEMATICS DEPARTMENT,
MOTILAL NEHRU NATIONAL INSTITUTE OF TECHNOLOGY
ALLAHABAD, U.P. (INDIA)

[3]MATHEMATICS DEPARTMENT,
UNIVERSITY OF ALLAHABAD, ALLAHABAD, (INDIA)

ABSTRACT. In the spirit of Diffie Hellman the concept of a protocol algebra is introduced using certain amalgamated free product of Braid group ($B$) and Thompson group ($T$) together with a nilpotent subgroup $H$ of index 2

## 1. INTRODUCTION

Most of the classical cryptographic schemes use Abelian groups in some way. In particular Diffie Hellman key exchange uses finite cyclic groups. So the term group based cryptography refers to cryptographic protocols that use infinite non Abelian group such as Braid groups. Braid groups can be used as a "platform" for a noncommutative cryptographic public key protocol. In this paper, in spirit of Diffe Hellman, a cryptosystem is generated using amalgamated free product of Braid groups and Thompson groups amalgamated through a subgroup $H$ whose commutator subgroup lies in the center of $H$.

**Definition 1.1.** *The Braid group on $n$ strands, denoted by $B_n$, is a group which has intuitive geometrical representation, and in a sense generalizesthe symmetric group $S_n$. The braid group $B_n$ on $n$ strands, is generated by $n-1$ generators $x_1, \ldots, x_{n-1}$ satisfying the following relations*

(1) $x_i x_j = x_j x_i$ *whenever* $|i - j| \geq 2$;
(2) $x_i x_{i+1} x_i = x_{i+1} x_i x_{i+1}$ *for* $i = 1, 2, \ldots, (n-2)$.

*Remark* 1.1.   (1) The groups $B_0$ and $B_1$ are trivial.
  (2) The group $B_2$ is generated by a single generator $x_1$ and non-empty set of relation. In general, if natural number $n > 1$, then $B_n$ is an infinite group.
  (3) The group $B_n$ for $n \geq 3$ is a nonabelian group.

$B_n$ is a subgroup $B_{n+1}$. It can be viewed as consisting of all those braid on $n + 1$ strands in which the bottom strand is horizontal and neither cross nor is crossed by any other strand. The simplest way to generalize the notion to an infinite number of strands is to take the direct limit of Braid groups, where the attaching maps $f : B_n \longrightarrow B_{n+1}$ send the $n-1$ generators of $B_n$ to the first $n-1$ generators of $B_{n+1}$ (i.e. by attaching a trivial strand). The *formal union of all the braid groups i.e.* $B = \bigcup_{i=1}^{\infty} B_i$ *is sometimes called the infinite group,* $B = \langle x_1, x_2, \ldots, x_i, \ldots | x_i x_j = x_j x_i$ *whenever* $|i - j| \geq 2$ *and* $x_i x_{i+1} x_i = x_{i+1} x_i x_{i+1}\rangle$.

**Definition 1.2.** *The Thompson Group* $T = \langle x_0, x_1, x_2, \ldots | x_k x_i = x_i x_{k+1} (k > i)\rangle$. *This presentation is infinite. There are also finite presentations of Thompson's group, for e.g.* $T = \langle x_0, x_1, x_2, x_3, x_4 | x_k x_i = x_i x_{k+1} (k > i, k < 4)\rangle$.

**Definition 1.3.** *If $G$ and $H$ are groups, a word in $G$ and $H$ is a product of the form $s_1 s_2 \ldots s_n$, where each $s_i$ is either an element of $G$ or an element of $H$. Such a word may be reduced using the following operations:*
- *Remove an instant of the identity element (of either $G$ or $H$)*
- *Replace a pair of the form $g_1 g_2$ by its product in $G$, or a pair $h_1 h_2$ by its product in $H$.*

*Every reduced word is an alternating product of elements of $G$ and $H$. For example:* $g_1 h_1 g_2 h_2 \ldots g_k h_k$. *The free product $G * H$ is the group whose elements are the reduced words in $G$ and $H$, under the operation of concatenation followed by reduction. The free product is always infinite. Suppose that $G = \langle R_G | S_G \rangle$ is a presentation for $G$, where $R_G$ is a set of generators and $S_G$ is a set of relations. Also $H = \langle R_H | S_H \rangle$ is a presentation for $H$, where $R_H$ is a set of generators and $S_H$ is a set of relations. Then $G * H = \langle R_G \bigcup R_H | S_G \bigcup S_H \rangle$ i.e $G * H$ is generated by the generators for $G$ together with the generators for $H$, with relations consisting of the relations from $G$ together with the relations from $H$ (assume here no notational clashes so that these are in fact disjoint union).*

**Example 1.4.** *Suppose that $G$ is a cyclic group of order 4 i.e. $G = \langle x | x^4 = 1\rangle$ and $H$ is a cyclic group of order 5 i.e. $H = \langle y | y^5 = 1\rangle$. Then $G * H = \langle x, y | x^4 = y^5 = 1\rangle$ is an infinite group.*

**Definition 1.5.** *Suppose $G$ has a presentation*
$\langle a_1, \ldots, a_n, b_1, \ldots, b_m | R(a_k), \ldots, S(b_l), \ldots, U_1(a_k) = V_1(b_l), \ldots, U_q(a_k) = V_q(b_l)\rangle$
*and we have*
(1) *$A$ is subgroup of $G$ generated by $a_1, a_2, \ldots, a_n$.*
(2) *$B$ is subgroup of $G$ generated by $b_1, b_2, \ldots, b_m$.*
(3) *$H$ is subgroup of $A$ generated by $U_1(a_k), \ldots, U_q(a_k)$, where $U_i(a_k)$ is a word in $a_1, a_2, \ldots, a_n$.*
(4) *$K$ is the subgroup of $B$ generated by $V_1(b_l), \ldots, V_q(b_l)$, where $V_j(b_j)$ is word in $b_1, b_2, \ldots, b_m$.*

*Then $G$ is called the free product of $A$ and $B$ with the subgroups $H$ and $K$ amalgamated under the mapping $U_i(a_k) \mapsto V_i(b_l)$.*

**Example 1.6.** *Consider $G = \langle a, b | a^4 = 1, b^6 = 1, a^2 = b^3\rangle$. The homomorphism of $G$ into $\langle x | x^{12} = 1\rangle$ given by $a \mapsto x^3, b \mapsto x^2$ shows that $a$ and $b$ have orders four and six respectively. Hence $G$ is the free product of $A$ and $B$ with the cyclic subgroups $H$ and $K$ of order two of $A$ and $B$ respectively amalgamated under the mapping $a^2 \mapsto b^3$, where $A = \langle a | a^4 = 1\rangle$ and $B = \langle b | b^6 = 1\rangle$.*

*Remark* 1.2. The free product of groups is a generalization of a free group; for a free group is the free product of infinite cyclic groups. Similarly, the free product of groups with an amalgamated subgroup is a generalization of the free product; for if the subgroup amalgamated is 1, then the free product results.

## 2. Fundamental Problems of Dehn

- **Word Problem**: Given a presentation $\langle X; R \rangle$ of a group $G$. For an arbitrary word $W$ in the generators, do we have an algorithm by which we can decide in a finite number of steps whether $W$ defines the identity element for $G$ or not.
- **Conjugacy Problem**: Given a presentation $\langle X; R \rangle$ of a group $G$. For two arbitrary words $W_1, W_2$ in the generators, do we have an algorithm by which we can decide in a finite number of steps whether $W_1$ and $W_2$ define conjugate elements of $G$ or not.

  The conjugacy problem is even more difficult than word problem.
- **Conjugacy Search Problem**: Given a presentation $\langle X; R \rangle$ of a group $G$ and the information that $W_1$ and $W_2$ are conjugate in $G$. DO we have an algorithm by which in a finite number of steps we can find a word $W_3$ such that $W_2 = W_3^{-1} W_1 W_3$.

## 3. Protocol

Consider braid group $B = \langle x_1, x_2, \ldots, x_i, \ldots | x_i x_j = x_j x_i$ whenever $|i-j| \geq 2$ and $x_i x_{i+1} x_i = x_{i+1} x_i x_{i+1} \rangle$ and Thompson group $T = \langle y_0, y_1, y_2, \ldots | y_k y_i = y_i y_{k+1} (k > i) \rangle$. Let $\{w_i | i \epsilon \lambda\}$ and $\{u_i | i \epsilon \lambda\}$ be set of words in $\{x_i\}$ and $\{y_i\}$ respectively. Let $H = \langle w_1, w_2, \ldots, w_n \rangle$ and $K = \langle u_1, u_2, \ldots, u_n \rangle$ be the subgroups of $B$ and $T$ respectively. Consider
$G = \langle x_1, x_2, \ldots, x_n, \ldots, y_0, y_1, \ldots | x_i x_j = x_j x_i$ whenever $|i-j| \geq 2$ and $x_i x_{i+1} x_i = x_{i+1} x_i x_{i+1}, y_k y_i = y_i y_{k+1} (k > i), w_1 = u_1, \ldots, w_n = u_n, w_i u_j w_i^{-1} u_j^{-1} w_l = w_l w_i u_j w_i^{-1} u_j^{-1} \rangle$ which is the amalgamated free product of $B$ and $T$ with subgroups $H$ and $K$ of $B$ and $T$ respectively. This is used as a platform group.

The group $G$ and $H$ & $K$ are made public.

- Sender computes $A = w_{i_1}^{\varepsilon_1} \ldots w_{i_L}^{\varepsilon_L}$, where $\varepsilon_k = \pm 1$ & $w_{i_k} \epsilon H$ and sends $(A^{-1} u_1 A, A^{-1} u_2 A, \ldots, A^{-1} u_n A)$ to receiver.
- Receiver computes $B = u_{j_1}^{\delta_1} \ldots u_{j_l}^{\delta_l}$, where $\delta_k = \pm 1$ & $u_{j_k} \epsilon K$ and sends $(B^{-1} w_1 B, \ldots, B^{-1} w_n B)$ to sender.
- Sender computes $K_1 = (A^{-1} B^{-1} w_1 BA, \ldots, A^{-1} B^{-1} w_n BA)$ and Receiver computes $K_2 = (B^{-1} A^{-1} u_1 AB, \ldots, B^{-1} A^{-1} u_n AB)$

  Since $B^{-1} A^{-1} u_i AB = A^{-1} B^{-1} (BAB^{-1} A^{-1}) u_i AB$
  $$= A^{-1} B^{-1} u_i (BAB^{-1} A^{-1}) AB$$
  $$= A^{-1} B^{-1} u_i BA$$
  $$= A^{-1} B^{-1} w_i BA \text{ (From definition of } G\text{)}$$
- Their secret key $K = K_1 = K_2$

To break, the protocol an adversary needs a solution to conjugacy search problem, because $K$ is conjugate to $(A^{-1} u_1 A, A^{-1} u_2 A, \ldots, A^{-1} u_n A)$ and $(B^{-1} w_1 B, \ldots, B^{-1} w_n B)$. Even if the presented group is known to be nilpotent group of class 2, the conjugacy search problem appears to be infeasible and therefore difficult for adversary to decrypt. For let $G$ be a nilpotent group of class 2. Suppose $g$ and $h$ are two conjugate elements i.e. there exist an element $u$ such that $g = u^{-1} h u = h h^{-1} u^{-1} h u$. Since $h^{-1} u^{-1} h u$ is an element of commutator and $G$ is a nilpotent group of class 2. So $g = h^{-1} u^{-1} h u h = (uh)^{-1} h u h$. Denote $v = uh$, then $g = v^{-1} h v$. This shows that there also exist an element of $G$ different from $u$ such that $g = v^{-1} h v$ and so on. Therefore the conjugacy search problem appears to be infeasible in $G$.

The conjugacy search problem in an amalgamated free product with a subgroup is more complicated even if the conjugacy search problem can be solved in $B$ and $T$ and the word problem can be solved in $G$. Thus the time-complexity increases in this protocol. It is still an open problem whether the conjugacy search problems in braid group can be solved in polynomial time by a deterministic algorithm.